\documentclass[aps,twocolumn,floats,superscriptaddress,prd,nofootinbib]{revtex4-1}
\usepackage{graphicx, epsfig, bm, amsmath}

\usepackage{color}
\usepackage{hyperref}
\usepackage{ wasysym }
\begin{document}

%%%%%%%%%%%%%%%%%%%%%%%%%%%%%%%%%%

\newcommand{\lcdm}{$\Lambda$CDM}

\newcommand{\gpr}{G^{\prime}}

\newcommand{\fnl}{f_{\rm NL}}
\newcommand{\curv}{{\cal R}}

\definecolor{darkgreen}{cmyk}{0.85,0.2,1.00,0.2}
\newcommand{\peter}[1]{\textcolor{red}{[{\bf PA}: #1]}}
\newcommand{\vin}[1]{\textcolor{darkgreen}{[{\bf VM}: #1]}}
\newcommand{\wh}[1]{\textcolor{blue}{[{\bf WH}: #1]}}

\newcommand{\WP}{W}
\newcommand{\XP}{X}
\newcommand{\B}{B^{\rm Bulk}}
\newcommand{\gB}{g_B}
\newcommand{\R}{\mathcal{R}}
\newcommand{\dotR}{\dot{\mathcal{R}}}
\newcommand{\ddotR}{\ddot{\mathcal{R}}}
\newcommand{\ep}{\epsilon_H}
\newcommand{\dotep}{\dot{\epsilon}_H}
\newcommand{\et}{\eta_H}
\newcommand{\dotet}{\dot{\eta}_H}
\newcommand{\cs}{c_s}
\newcommand{\esq}{\left(}  % esq = esquerda = left
\newcommand{\dir}{\right)} % dir = direita  = right
\newcommand{\ord}{\mathcal{O}}
\newcommand{\hR}{\hat{\mathcal{R}}}
\newcommand{\dothR}{\dot{\hat{\mathcal{R}}}}

\newcommand{\aap}{Astron. Astrophys.}

%%%%%%%%%%%%%%%%%%%%%%%%%%%%%%%%%%%%%%%%%%%

\pagestyle{plain}

\title{Warp Features in DBI Inflation}

\author{  Vin\'icius Miranda}
\affiliation{Department of Astronomy \& Astrophysics, University of Chicago, Chicago IL 60637}
\affiliation{The Capes Foundation, Ministry of Education of Brazil, Bras\'ilia DF 70359-970, Brazil}

\author{ Wayne Hu}
\affiliation{Department of Astronomy \& Astrophysics, University of Chicago, Chicago IL 60637}
\affiliation{Kavli Institute for Cosmological Physics,  Enrico Fermi Institute, University of Chicago, Chicago, IL 60637}

\author{Peter Adshead}
\affiliation{Kavli Institute for Cosmological Physics,  Enrico Fermi Institute, University of Chicago, Chicago, IL 60637}

\begin{abstract}
In Dirac-Born-Infeld inflation,  changes in the sound speed that transiently break the slow roll approximation lead to features in the power spectrum.
  We develop and test the generalized slow roll approximation for calculating such effects and show that it can be extended to treat order unity features.   As in slow-roll, model independent constraints on the potential of canonical inflation can be directly reinterpreted in the DBI context through this approximation.   In particular, a sharp horizon scale step in the
  warped brane tension can explain oscillatory features in the WMAP7 CMB power spectrum as well as features in the potential.   Differences appear only as a small suppression of
 power on horizon scales and larger.
  \end{abstract}

\maketitle

%=================================================================
\section{Introduction}
\label{sec:intro}

In Dirac-Born-Infeld (DBI) inflation \cite{Silverstein:2003hf,Alishahiha:2004eh}, transient but rapid changes in the sound speed  leave their imprint as features on the power spectrum.  For string-motivated DBI examples, such features might arise from duality cascades which impart steps in the warped brane tension 
\cite{Hailu:2006uj,Bean:2008na}. Annihilation of branes during DBI inflation has also been shown to lead to particle production and to the imprint of features on the warp \cite{Firouzjahi:2010ga}. More generally, within the context of effective field theory \cite{Cheung:2007st} it has been shown that a sharp step in the sound speed leads to oscillatory features in the power spectrum of fluctuations \cite{Park:2012rh}.

Power spectrum features from sudden changes in the warped brane tension of DBI inflation are
closely related to those from sudden changes in the potential 
for canonical single field inflation.    Measurements of the CMB temperature power spectrum from WMAP place observational constraints on the latter.  
Recently, the generalized slow roll approach (GSR) \cite{Stewart:2001cd,Dvorkin:2009ne} has been used to extract model-independent constraints  from the WMAP data on features as sharp as 1/4 of an efold \cite{Dvorkin:2010dn,Dvorkin:2011ui}.   Even sharper features lead to highly oscillatory power spectrum features which can evade these constraints
due to projection effects.  Indeed there is a special case where a sharp step in the potential  on scales near the current horizon can fit the WMAP data better that a smooth model in the acoustic regime
\cite{Adshead:2011jq}.  

The GSR approach remains valid for single field inflation with non-canonical kinetic terms
\cite{ArmendarizPicon:1999rj}, including DBI inflation, with a suitable reinterpretation of
the source of deviations from slow-roll \cite{Hu:2011vr}.  In this Paper, we develop the GSR approach for DBI inflation and show how observational constraints on potential features
translate to constraints on  warp features.

In \S \ref{sec:DBI}, we briefly review the phenomenology of DBI inflation and the exact
computation of its power spectrum.   In \S \ref{sec:GSR} we develop and test the GSR approach in the DBI context and establish the correspondence between potential features and warp features.  In \S \ref{sec:step}, we consider the special
case of a sharp step in the warp analytically and show that it can explain the WMAP data 
as well as a sharp step  in the potential.   We discuss these results in \S \ref{sec:discussion}.

\section{DBI Inflation}
\label{sec:DBI}

We consider DBI inflation to be a phenomenological model with the Lagrangian density
\begin{equation}
{\cal L}= \left[
1-\sqrt{1 - 2  X/T(\phi)} \right] T(\phi)- V(\phi),
\end{equation}
where the kinetic term $2X = - \nabla^{\mu} \phi \nabla_{\mu} \phi$.  
In braneworld theories that motivate the DBI Lagrangian, $\phi$ determines the position of the brane, $T(\phi)$ gives the warped brane tension, and $V(\phi)$ is the interaction potential.  

As a consequence of the non-canonical kinetic structure, field perturbations 
propagate at the sound speed
\begin{equation}
c_s(\phi,X) =\sqrt{ 1 - 2 X/T(\phi)}.
\end{equation}
The inflaton energy density and pressure can be expressed in terms of the sound
speed as
\begin{eqnarray}
\rho(\phi,X)  &=& \left(  \frac{1}{c_s} -1 \right)T(\phi)  + V(\phi), \nonumber\\
p(\phi,X) &=&  ({1-c_s})T(\phi) - V(\phi) .
\label{eqn:energypressure}
\end{eqnarray}
Note that for $X /T \ll 1$, $c_s=1$ and the Lagrangian, $\rho$ and $p$ take on their canonical forms.

For the background equations of motion, we take the acceleration equation
\begin{equation}
\frac{H_N}{H} = - \frac{\phi^2_N}{2 c_s} \equiv -\epsilon_H,
\label{eqn:acceleration}
\end{equation}
where $X= H^2 \phi_N^2/2$, the Hubble parameter satisfies the Friedmann equation $H^2=\rho/3$, 
and the field equation
\begin{eqnarray} \label{eqn::eom_phi}
\phi_{NN}& =& - \left( \frac{H_N}{H} + 3 c_s^2\right)  \phi_N - c_s^3 \frac{V_\phi}{H^2} \nonumber\\
&&\quad + \frac{1}{2}  (1-c_s)^2(1+2 c_s)\frac{T_\phi}{H^2}.
\label{eqn:field}
\end{eqnarray}
Here and throughout the subscript $N$ denotes $d/d\ln a$,
the subscript $\phi$ likewise $d/d\phi$, and we choose units where
$M_{\rm pl}=(8\pi G)^{-1/2} = c = \hbar = 1$.
If warp and potential features are absent near the initial conditions,
initial values for $\{ \phi, \phi_N,H \}$ can be set on the slow-roll attractor  
\begin{equation}
\phi_N \approx -\frac{c_s}{3} \frac{V_\phi}{H^2}  , \quad H^2 \approx \frac{V}{3},
\label{eqn:attractor}
\end{equation}
where we assume that the $V_\phi$ term dominates over $T_\phi$.
Given that we choose to solve Eq.~(\ref{eqn:acceleration}) and (\ref{eqn:field}), we must
ensure that the Friedmann equation is exactly satisfied on the initial condition
\cite{Mortonson:2009qv}.  
This can be achieved by first choosing the initial $\phi(N_i)$, then taking
\begin{equation}
H \phi_N \big|_{N_i}  = -\sqrt{ \frac{V}{3}} \frac{V_\phi}{V} c_s , \quad
c_s \big|_{N_i}=
\left( 1+ \frac{V}{3 T}\frac{ V_\phi^2}{V^2}  \right)^{-1/2}, 
%  \left( 1 + \frac{V}{3 T}\frac{ V_\phi^2}{V^2} \right)^{-1/2},
\label{eqn:HphiN}
\end{equation}
and calculating $\rho$, $H$ exactly through Eq.~(\ref{eqn:energypressure}).   Since $\phi_N= H\phi_N/H$, we now have a self-consistent set of initial conditions $\{ \phi,\phi_N,H \}$ at $N_i$.
This technique remains valid for all $c_s$ in the slow-roll approximation.   On the other hand, for $c_s \ll 1$ the  slow-roll approximation can remain valid even for
steep potentials in Eq.~(\ref{eqn:attractor}).

We evolve these equations until the field reaches $\phi=\phi_{\rm end}$ which we take to be the end of inflation and define  $N = 0$ to be this epoch
\begin{equation}
N = \int_{\ln a}^{\ln a_{\rm end}} d\ln a,
\end{equation}
such that $N<0$ during inflation.
For the purposes of calculating the power spectrum, it is useful to express
the efolding number, $N$, in terms of the sound horizon, the comoving distance sound can travel from $N$ to the end of 
inflation
\begin{eqnarray}\label{eqn:def_sound_horizon}
	s(N) = \int_{N}^{0} d \tilde{N} {c_s \over aH}  = \frac{1}{a_{\text{end}}}  \int_{N}^{0} d\tilde{N} {c_s \over e^{\tilde{N}} H } .
\end{eqnarray}
By defining the effective reheat temperature as $T_{\text{reheat}} \equiv T_0/a_\text{end}$, where the present CMB temperature is $T_0=2.726$K, we can
express the sound horizon as
\begin{equation}
	\frac{s(N)}{500 \text{Mpc} } = e^{-65.08} T_{\text{reheat}} \int_{N}^{0} d \tilde{N} {c_s \over aH}.
\end{equation}
The curvature power spectrum is then given by 
\begin{align}\label{eqn:power_spectrum}
        \Delta_{\R}^2 \equiv \frac{k^3 P_{\R}}{2 \pi^2} %        
        = \lim_{ks \to 0} \left| \frac{k s y}{f} \right|^2 ,
\end{align}
where the mode function $y$ obeys the Mukhanov-Sasaki equation
\cite{Garriga:1999vw,Hu:2011vr}
\begin{align}\label{eqn:yeqn}
        \frac{d^2y}{ds^2} + \left(k^2 - \frac{2}{s^2} \right) y = \frac{g(\ln s)}{s^2}y.
\end{align}
Here
\begin{align}
        g \equiv \frac{f'' - 3 f'}{f},
\end{align}
with  $' \equiv d/d\ln s$ throughout and
\begin{align}\label{eqn:fdef}
	f^2 & = 8 \pi^2 \frac{\ep \cs}{H^2} \esq \frac{a H s}{\cs} \dir^2.
\end{align}
We solve Eq.~(\ref{eqn:yeqn}) assuming Bunch-Davies initial conditions where $\lim_{ks \gg 1} y = e^{iks}$.
Note that written in this form, the Mukhanov-Sasaki equation takes exactly
the same form for canonical and non-canonical kinetic terms.  This fact will allow
us to remap existing constraints on $g(\ln s)$ onto DBI models.

\section{GSR Approximation}
\label{sec:GSR}

In this section, we begin by reviewing the GSR approach to calculating the curvature
power spectrum
\cite{Stewart:2001cd,Dvorkin:2009ne,Hu:2011vr}
and show how to apply it to the DBI model.  We then test the
accuracy of the approach against the exact computation for a step-like feature
in the warped brane tension \cite{Bean:2008na} and show that the model-independent constraints on features in the potential for canonical single field inflation
\cite{Dvorkin:2010dn,Dvorkin:2011ui}
 can be readily reinterpreted in the DBI context. 

\subsection{Technique}

Briefly,
the GSR approach to solving the Mukhanov-Sasaki equation (\ref{eqn:yeqn})
 is to consider the RHS as an external source with an iterative correction to the
 field value $y$.    To lowest order, we replace $y \rightarrow y_0$ where 
\begin{equation}
y_0 = \left( 1 + {i \over ks} \right) e^{i ks} ,
\end{equation}
is the solution to equations with $g \rightarrow 0$ and solve for the field fluctuation $y$ through the Green function technique.   To second order in slow-roll, the curvature power spectrum is given by \cite{Choe:2004zg,Dvorkin:2010dn}
\begin{align}
\label{eqn:GSRpower}
\ln \Delta_\curv^{2} &= G(\ln s_{\rm min}) + \int_{s_{\rm min}}^\infty {d s\over s} W(ks) G'(\ln s)\\
&\quad + \ln \left\{ [ 1+ {1\over 4}I_1^2(k) + {1\over 2}I_2(k)]^2 + {1 \over 2}I_1^2(k) \right\} , \nonumber
\end{align}
where 
 the window function
\begin{equation}
W(u) = {3 \sin(2 u) \over 2 u^3} - {3 \cos (2 u) \over u^2} - {3 \sin(2 u)\over 2 u} .
\end{equation}
Here
\begin{equation}
G = - 2 \ln f    + {2 \over 3} (\ln f )'   ,
\end{equation}
and thus 
\begin{equation}
G' = -2 (\ln f )' + {2 \over 3}  (\ln f )''  = {2 \over 3}  g - {2 \over 3} [(\ln f)']^2  .
\label{eqn:Gprime}
\end{equation}
We  call $G'$ the GSR source function.  The quadratic term in $(\ln f)'$ appears
to ensure constant curvature fluctuations above the sound horizon.

The $I_1$ and $I_2$ integrals are the second order corrections
\begin{eqnarray}
I_1(k) &=& { 1\over \sqrt{2} } \int_0^\infty {d s \over s} G'(\ln s) X(ks) , \nonumber\\
I_2(k) &=& -4 \int_0^\infty { d u \over u } [ X + {1\over 3} X' ] {f' \over f} F_2(u) ,
\end{eqnarray}
with $u=k s$,
\begin{equation}
F_2(u) = \int_u^\infty {d \tilde u \over \tilde u^2} {f' \over f},
\end{equation}
and
\begin{equation}
X(u) = {3 \over u^3} (\sin u - u \cos u)^2 .
\end{equation}

To calculate the power spectrum in the GSR approximation, we need to obtain the
source functions $G'$ and $f'/f$ from the solution to the background equations of motion
(\ref{eqn:acceleration}) and (\ref{eqn:field}).  In terms of the slow-roll parameters
\cite{Hu:2011vr}
\begin{eqnarray} \label{eqn:dG}
G'& =& {2 \over 3} (2 \epsilon_H -2\eta_H -\sigma_1) + 
{2 \over 3} ({ a H s \over c_s }-1)^2 \\
&& +  {2 \over 3} ({ a H s \over c_s} -1) (4 + 2\epsilon_H - 2\eta_H - \sigma_1) \nonumber\\
&& +{1 \over 3}  \left( { a H s \over c_s}\right)^2 
\Big[ 2 \delta_2 + 2\epsilon_H^2 - 2\eta_H - 2\eta_H^2 
\nonumber\\
&& - 3\sigma_1 + 2 \eta_H \sigma_1 
+ \sigma_1^2 - \epsilon_H ( 4\eta_H+\sigma_1) - \sigma_2\Big], \nonumber
\end{eqnarray}
and
\begin{equation}
\frac{f'}{f}= \left( \frac{ a H s }{c_s}\right)  \left(\eta_H-\epsilon_H+\frac{1}{2}\sigma_1\right) +
\left( 1 -  \frac{ a H s }{c_s}\right)
\end{equation}
where the additional slow-roll parameters are defined by
\begin{align}\label{eqn:epsilon_def}
		 \eta_H &\equiv \epsilon_H - \frac{1}{2} \frac{d\ln\epsilon_H}{dN}, \nonumber\\
		\delta_2 &\equiv \epsilon_H \eta_H + \eta_H^2 - \frac{d\eta_H}{dN}, \nonumber\\
		\sigma_1 & \equiv \frac{d\ln c_s}{dN} ,\nonumber\\
		\sigma_2 & \equiv \frac{d\sigma_1}{dN} .
\end{align}      

Using the field equation (\ref{eqn:field}), we can write
\begin{align}
	\frac{\phi_{NN}}{\phi_N} = \epsilon_H - c_s^2 \tilde{\eta}_H + \frac{1}{2}   \frac{(1-c_s)(1+2c_s)}{1+c_s}\tilde{\sigma}_1,
\end{align}
where
\begin{align}
\tilde \eta_H & \equiv  \left( 3 + \frac{V_\phi c_s}{\phi_N H^2} \right), \nonumber\\
\tilde \sigma_1 &\equiv \frac{T_\phi}{T}\phi_N.
\end{align}
These auxiliary parameters $\tilde\eta_H$ and $\tilde \sigma_1$ quantify  slow-roll deviations generated by features in the potential $V_\phi$ and features in the warp $T_\phi$
respectively.
 
In terms of the auxiliary parameters, the slow roll parameters themselves become
\begin{align}
\eta_H  &=
		 \frac{1+c_s^2}{2} \tilde \eta_H-\frac{c_s}{2}  \frac{1-c_s}{1+c_s}\tilde\sigma_1,  \nonumber\\
\sigma_1 &=  (1-c_s) \tilde\sigma_1 + (1 - c_s^2)\tilde\eta_H .
\label{eqn:etahsigma1}
\end{align} 
Note that for $\eta_H$, the term involving $\tilde \sigma_1$ is suppressed both as $c_s\rightarrow0$ and
$c_s\rightarrow 1$.
Furthermore $\tilde \eta_H$ is slow roll suppressed on the attractor of Eq.~(\ref{eqn:attractor}) and for $c_s=1$, 
$\eta_H=\tilde\eta_H$.     If features in $T(\phi)$ drive deviations from slow-roll then the
$\tilde \sigma_1$ term dominates, $\eta_H =[ c_s/2(1+c_s)] \sigma_1$ and hence
$|\eta_H| \ll |\sigma_1|$ for $c_s\ll 1$.

The remaining slow roll parameters $\sigma_2$ and $\delta_2$ can be constructed
by taking the derivatives of $\sigma_1$ and $\eta_H$
\begin{align}
\sigma_2 &= (1-c_s)\frac {d\tilde \sigma_1}{dN} - c_s \sigma_1 \tilde \sigma_1
+ (1-c_s^2) \frac{d\tilde \eta_H}{dN}- 2 c_s^2\sigma_1 \tilde\eta_H, \nonumber\\
%\frac{d\eta_H}{dN} &= \frac{1+c_s^2}{2} \frac{d\tilde \eta_H}{dN} -  c_s^2\sigma_1\tilde \eta_H  \nonumber\\
%&\quad -\frac{c_s }{2} \frac{1-2c_s-c_s^2}{(1+c_s)^2}   \sigma_1 \tilde\sigma_1 
%+ \frac{c_s}{2}  \frac{1-c_s}{1+c_s}\frac{d \tilde\sigma_1}{dN} \nonumber\\
\delta_2 &= - \frac{1+c_s^2}{2} \frac{d\tilde \eta_H}{dN}  + \epsilon_H\eta_H + \eta_H^2 - c_s^2\sigma_1\tilde \eta_H  \nonumber\\
&\quad + \frac{c_s}{2}  \frac{1-c_s}{1+c_s}\frac{d \tilde\sigma_1}{dN}+\frac{c_s }{2} \frac{1-2c_s-c_s^2}{(1+c_s)^2}   \sigma_1 \tilde\sigma_1 ,
\end{align}
where
\begin{align}
\frac{d \tilde \eta_H}{dN} & = c_s \frac{V_{\phi\phi}}{H^2} + \frac{c_s V_\phi}{\phi_N H^2}
\left( \sigma_1 - \frac{\phi_{NN}}{\phi_N} + 2\epsilon_H \right) \nonumber\\
& =  c_s \frac{V_{\phi\phi}}{H^2} + \left({\tilde\eta_H} - 3\right) \left( \epsilon_H + \tilde{\eta}_H + \frac{1 - c_s}{1 + c_s} \frac{\tilde\sigma_1}{2} \right), \nonumber \\
\label{eqn:VTphiphi}
\frac{d \tilde \sigma_1}{dN} &= \frac{T_{\phi\phi}}{T} \phi_N^2 - \left( \frac{T_\phi}{T} \phi_N\right)^2 + \frac{T_\phi}{T}\phi_{NN}\\
&= \frac{T_{\phi\phi}}{T} \phi_N^2 -\frac{(1 + c_s + 2 c_s^2)}{2(1+c_s)}\tilde{\sigma}_1^2 +  (\epsilon_H - c_s^2 \tilde{\eta}_H)\tilde{\sigma}_1 .\nonumber 
\end{align}
For sharp features in the warp and potential, $\sigma_2$ and $\delta_2$ dominate
respectively in $G'$ due to the appearance of second derivatives in Eq.~(\ref{eqn:VTphiphi}).
 Note that since $\phi_N^2 = 2\epsilon_H c_s$, the impact of
fractional features in the warp vs.\ the potential potential is suppressed by slow roll parameters.

In the slow roll limit, one can iteratively substitute the attractor solution Eq.~(\ref{eqn:attractor}) into the field equation to obtain
\begin{align}\label{eqn:gprimeapprox}
G' & \approx 4\epsilon_H -2 \eta_H + \sigma_1 \\
& \approx c_s (2+c_s^2) \left( {\frac{V_\phi}{V}}\right)^2 - 2 c_s^3 \frac{V_{\phi\phi}}{V} 
- c_s (1-c_s^2) \frac{T_\phi}{T}\frac{V_\phi}{V} ,  \nonumber
\end{align}
where $c_s(\phi)$ is given by the attractor solution Eq.~(\ref{eqn:HphiN}).
The absence of a $T_{\phi\phi}$ term in Eq.\ (\ref{eqn:gprimeapprox}) can be attributed to the fact that the attractor solution is determined by $V_\phi$. 
Furthermore, in the slow roll limit, evolution in $G'$ is second order in slow roll parameters and
\begin{equation}
\frac{f'}{f} \approx -2\epsilon_H+\eta_H-\frac{1}{2}\sigma_1 \approx -\frac{1}{2}G'
\end{equation} 
and so
\begin{align}
        I_1  & \approx \frac{\pi}{2 \sqrt{2}} (4\epsilon_H -2 \eta_H + \sigma_1)    ,\nonumber\\
            I_2 & \approx  -4 \left(\frac{f'}{f}\right)^2 \nonumber\\
        &\approx - \left( 4\epsilon_H -2 \eta_H + \sigma_1\right)^2 .
        \label{eqn:Islowroll}
\end{align}
Thus in the slow roll limit the total second order correction involves a near
cancellation of the $I_1$ and $I_2$ terms
\begin{align}
& \ln \left\{ [ 1+ {1\over 4}I_1^2(k) + {1\over 2}I_2(k)]^2 + {1 \over 2}I_1^2(k) \right\}  \approx I_1^2 + I_2 \nonumber\\
& \quad \approx \left( \frac{\pi^2}{8} - 1 \right) \left( 4\epsilon_H -2 \eta_H + \sigma_1\right)^2.
\end{align}

  If sharp features in either the potential or the warp dominate
\begin{align}
G' &\approx \frac{2}{3}\delta_2 - \frac{1}{3}\sigma_2 \nonumber\\
&\approx -{2} c_s \frac{V_{\phi\phi}}{V} -\frac{1-c_s}{3(1+c_s)}\frac{T_{\phi\phi}}{T}\phi_N^2 .
\end{align}
Unlike the canonical case, the functional constraints imposed by observational constraints on $V(\phi)$ and
$T(\phi)$ differ in the two limits.   Nonetheless deviations in the power spectrum
 for the sharp feature case, which  can be large yet observationally viable, 
share strong similarities between those generated by
$T(\phi)$ and $V(\phi)$.  Also, unlike the slow-roll limit, $I_1$ tends to be larger than $I_2$ in that only it depends directly on second derivatives of $T(\phi)$ or
$V(\phi)$ as we shall see explicitly in the next section.

\subsection{Numerical Tests}

The GSR construction in the previous section applies to any model with features in the warped brane tension
$T(\phi)$ or potential $V(\phi)$.
For definiteness and motivated by the WMAP data, we test the GSR approximation on models where $T(\phi)$ has a step feature \cite{Bean:2008na}
\begin{align}\label{eqn:T(phi)}
	T(\phi) = \frac{\phi^4}{\lambda_B} [1 + b F(\phi) ], 
\end{align}
with
\begin{align}
	  F(\phi) = \tanh\left(\frac{\phi - \phi_s}{d} \right)-1 
\end{align}
and $\phi$ inflates on a potential
\begin{equation}
V(\phi) =  V_0(1 - \frac{1}{6}\beta \phi^2),
\end{equation}
rolling from small to large values.  We have chosen a convention that after the feature, $T(\phi)$ goes back to its $b=0$ value.   Since physical
scales are matched to the end of inflation through Eq.~(\ref{eqn:def_sound_horizon}), this
simplifies the comparison to the smooth featureless case.
For simplicity, we will take $T_{\rm reheat}=V_0^{1/4}$ following \cite{Bean:2007eh}.
Thus the DBI step model is  specified by 4 parameters \{$\lambda_B, V_0, \beta, \phi_\text{end}$\}
controlling the underlying smooth spectrum and 3 parameters describing the step feature
\{$\phi_s$,$b$,$d$\}.   

In order to set the parameters for the smooth $b=0$ spectrum, it is useful to re-express the attractor
solution of Eq.~(\ref{eqn:attractor}) in terms of efolds \cite{Bean:2008na}
\begin{equation}
\phi_N = \frac{d\phi}{dN} \approx \frac{\phi^2}{H\sqrt{\lambda_B}},
\end{equation}
and hence
\begin{align}
\phi(N) & \approx -  H\sqrt{\lambda_B}\frac{1}{N-N_0}, \nonumber\\
c_s(N) & \approx -\frac{3}{\beta}\frac{1}{N-N_0}, \nonumber\\
\epsilon_H(N) &\approx -\frac{\beta}{6} H^2\lambda_B \frac{1}{(N-N_0)^3} ,
\end{align}
where $N_0$ is an integration constant determined by our definition that
 $N(\phi_{\rm end})=0$.  In the slow roll approximation
 \begin{align}
 \Delta_\curv^2 &\approx \left( { H \over 2\pi \phi_N}\right)^2 = \frac{H^2}{8\pi^2\epsilon_H c_s}
 \nonumber\\
 & \approx \frac{(N-N_0)^4}{4\pi^2 \lambda_B},
 \label{eqn:Pksr}
 \end{align}
 and the tilt
 \begin{align}
 n_s -1 \equiv \frac{d \ln \Delta_\curv^2}{d\ln k} \approx \frac{4}{N-N_0} .
 \end{align}
Note that to have a tilt that is compatible with observations $n_s -1 \sim -0.04$ at 
$N \sim -50$ one requires $N_0 \sim 50$.   In string-inspired models where
$\phi_{\rm end} \approx  H \sqrt{\lambda_B}$, $N_0 = {\cal O}(1)$.  
While such problems can be ameliorated by introducing stringy physics not included
in the DBI action \cite{Bean:2007eh}
doing so degrades the predictive power of calculations based on this action
({\it cf.}
 \cite{Bean:2008na}).
We therefore instead
require $\phi_{\rm end} \ll H\sqrt{\lambda_B}$ so that inflation ends while the field is
deep in the DBI regime.     Finally, to satisfy constraints from upper limits on 
equilateral type bispectra, we require $c_s >1/30$ for scales relevant to the CMB \cite{Komatsu:2010fb}.

These conditions are satisfied by the following choices
\begin{align}
\lambda_B &= 1.93 \times 10^{15}, \nonumber\\
V_0 &= 7.10 \times 10^{-26} , \nonumber\\
\beta &= 0.5, \nonumber\\
\phi_{\rm end} &= 1.065 \times 10^{-7}.
\end{align}
In our parameterization $V_0$ drops out of expressions for the curvature power spectrum at a fixed $N-N_0$ or $n_s-1$ and only impacts the mapping between field and physical scale through $T_{\rm reheat}$.   It also enters into the tensor-scalar ratio and so we take for definiteness a small
value such that tensors are negligible.

Since the GSR approximation reproduces the exact second order expansion in
slow roll parameters by construction when they are all small, we test the technique  for the nontrivial case
where $b$ is order unity.

\begin{figure}[t]
\psfig{file=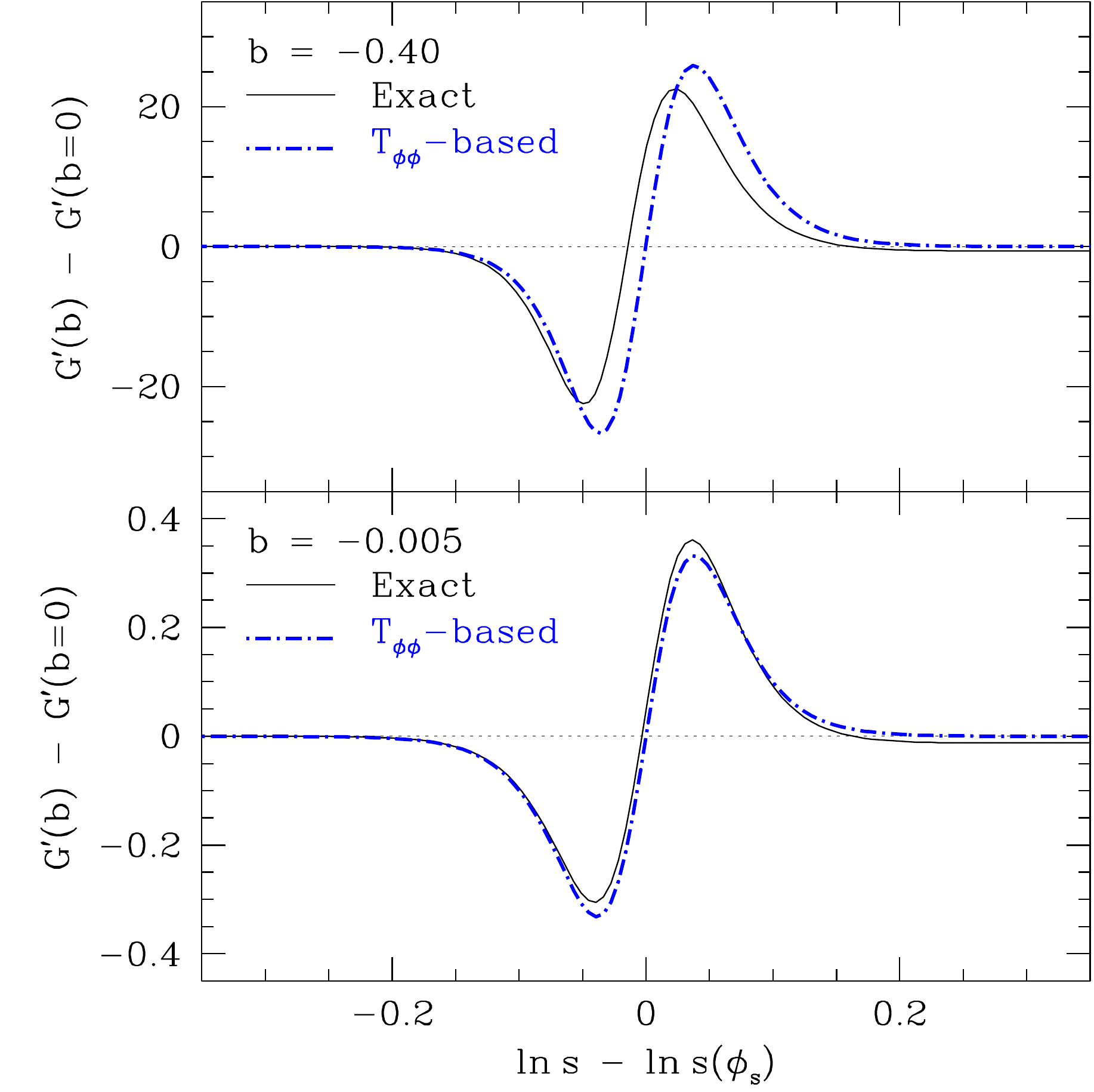, width=3.25in}
\caption{GSR source function $G'$ for a warp step with $b= -0.4$ (top) and
$b=-0.005$ (bottom) with $d = 2.81 \times 10^{-11}$    Also shown is the approximate form based on $T_{\phi\phi}$ in the sharp, small amplitude limit from Eq.~\eqref{eqn:appdG} which is an excellent approximation for the small $b$ case and remains in good qualitative agreement for the high $b$ case.   }
\label{fig:dG}
\end{figure}

In Fig.~\ref{fig:dG}, we show an example of the GSR source function $G'$
where $b=-0.4$ (top) and $-0.005$ (bottom) and $d= 2.81 \times 10^{-11}$ and
$\phi_s = 5.67 \times 10^{-8}$.   Both cases appear like the second derivative
of the step in $T(\phi)$ with a width determined by the number of efolds it takes for
the inflaton to cross the step
\begin{equation}
\delta \ln s \approx \delta N \approx \frac{d}{\phi_N}.
\end{equation}
The main difference is that at the larger $b$ value
the location, amplitude and width of the feature differ slightly.

\begin{figure}[t]
\psfig{file=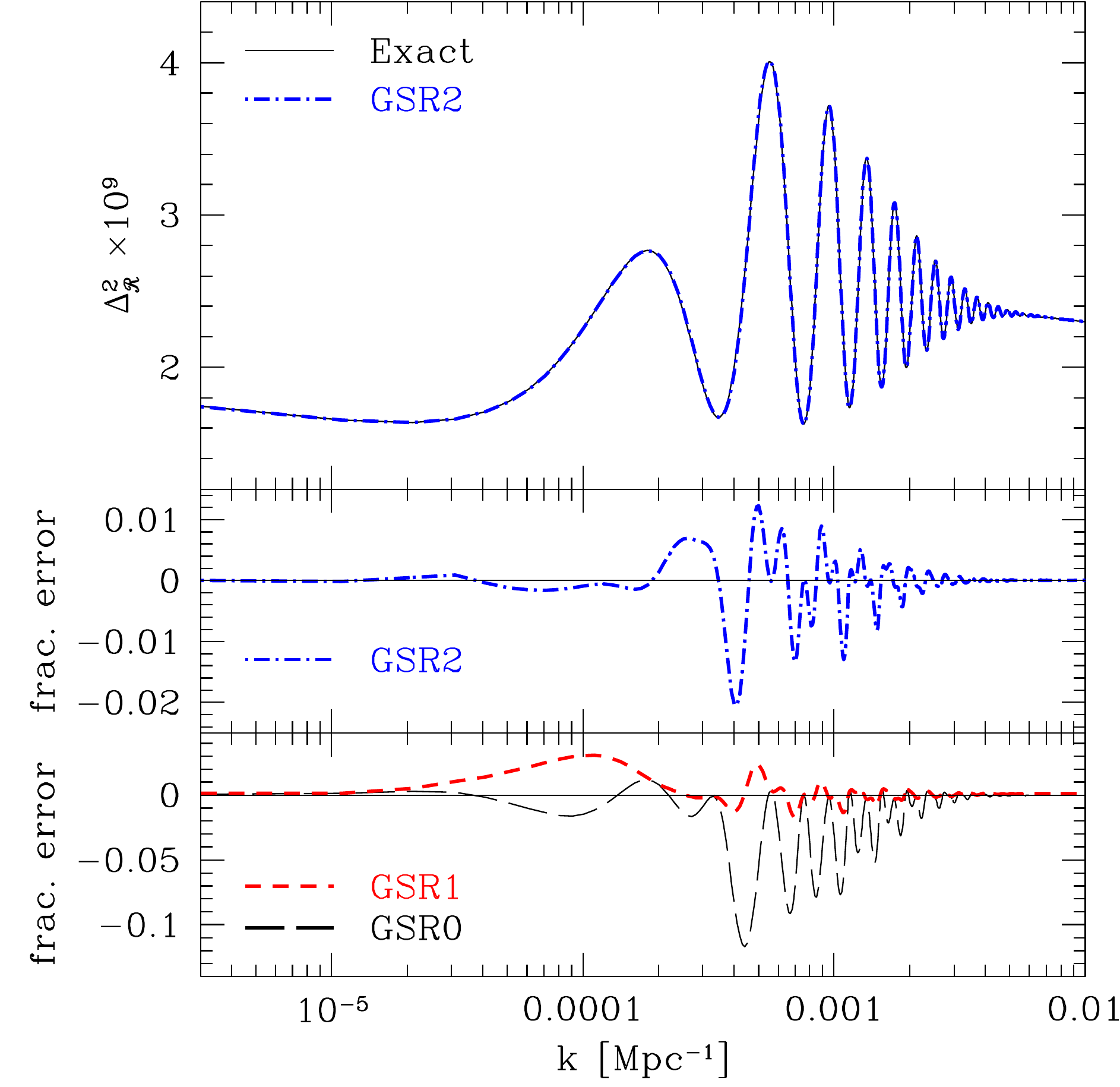, width=3.25in}
\caption{GSR vs exact solution for the power spectrum (top panel), the fractional difference between the two (middle panel), and the impact of second order corrections (bottom panel) corresponding to the $b=-0.4$ model in Fig.~\ref{fig:dG} (top).   GSR2 denotes the full solution (\ref{eqn:GSRpower})
with $I_1$ and $I_2$, GSR1 the solution with $I_2=0$ and GSR0 with $I_1=I_2=0$.   While the GSR2 solution captures effects at the $1-2\%$ level for $b=-0.4$, even the leading order GSR0 is accurate at the $10\%$ level.}
\label{fig:GSR2}
\end{figure}

In Fig.~\ref{fig:GSR2} we show the corresponding power spectrum.
In the top panel, we compare the power spectrum from the full GSR approximation
(``GSR2") of Eq.~(\ref{eqn:GSRpower}) to the exact solution.  In the middle panel, we show that the approximation 
is accurate at the 1-2\% level for the order unity feature.
Moreover, the second order corrections remain small as shown in the bottom panel
where ``GSR1" denotes setting $I_2=0$ and ``GSR0" denotes setting both $I_1=I_2=0$
in  Eq.~(\ref{eqn:GSRpower}).   Here the maximum value that $|I_1|$ attains is 0.37.
  As in the 
canonical case, ${\rm max}|I_{1}|<1/\sqrt{2}$ ensures accuracy in the power spectrum of the GSR approximation, typically to a few percent in observables such as the CMB power spectrum \cite{Dvorkin:2011ui}.   

Note $I_2$ provides a negligible absolute correction for
order unity and smaller features.   For small features $I_1$ and $I_2$ corrections do become comparable but in that case both are negligible (see Eq.~\ref{eqn:Islowroll}).    Since both leading order and $I_1$ terms depend only on a single source function $G'(\ln s)$, observational constraints from the power spectrum may be directly mapped onto constraints on this GSR source function
\cite{Dvorkin:2009ne}.

\subsection{Observational Constraints on Broad Features}

Model-independent analysis
of features in the source function $G'$ have been conducted using a principal
component basis that is complete for the WMAP7 data set for broad inflaton 
features that are traversed in $\delta N > 1/4$
\cite{Dvorkin:2010dn,Dvorkin:2011ui}.  In the acoustic regime of
$s \sim 100-400$ Mpc, constraints on possible deviations are extremely tight with percent level limits on the broadest features \cite{Dvorkin:2010dn}.   
As illustrated in Fig.~\ref{fig:dG}, these constraints can be interpreted in the DBI context as limits on
\begin{equation} \label{eqn:appdG}
G' \approx -\frac{2(1-c_s)}{3(1+c_s)}c_s \epsilon_H \frac{T_{\phi\phi}}{T}.
\end{equation}
 The only broadband feature that marginally improves the likelihood is associated with the known
$\ell \sim 20-40$ glitch in the WMAP data.  However, a simple step in the warp does not fit the data as well as a step in the potential due to the change in
$c_s$ across the step \cite{Bean:2008na}.  Since the attractor solution for the
power spectrum in Eq.~(\ref{eqn:Pksr}) depends on the sound speed, a step leads not only to oscillations, but also a step in the power spectrum
across the feature (see Fig.~\ref{fig:GSR2}).

\section{Sharp Step}
\label{sec:step}

Sharp features in the warped brane tension that are traversed by the inflaton in
$\delta N \ll 1$ produce high frequency oscillations in the
curvature power spectrum.  These are more difficult to constrain observationally than
broadband features due to projection effects and sky coverage.  
They are also significantly more cumbersome to calculate
as their effects persist over orders of magnitude in wavenumber.  

\begin{figure*}[t]
\psfig{file=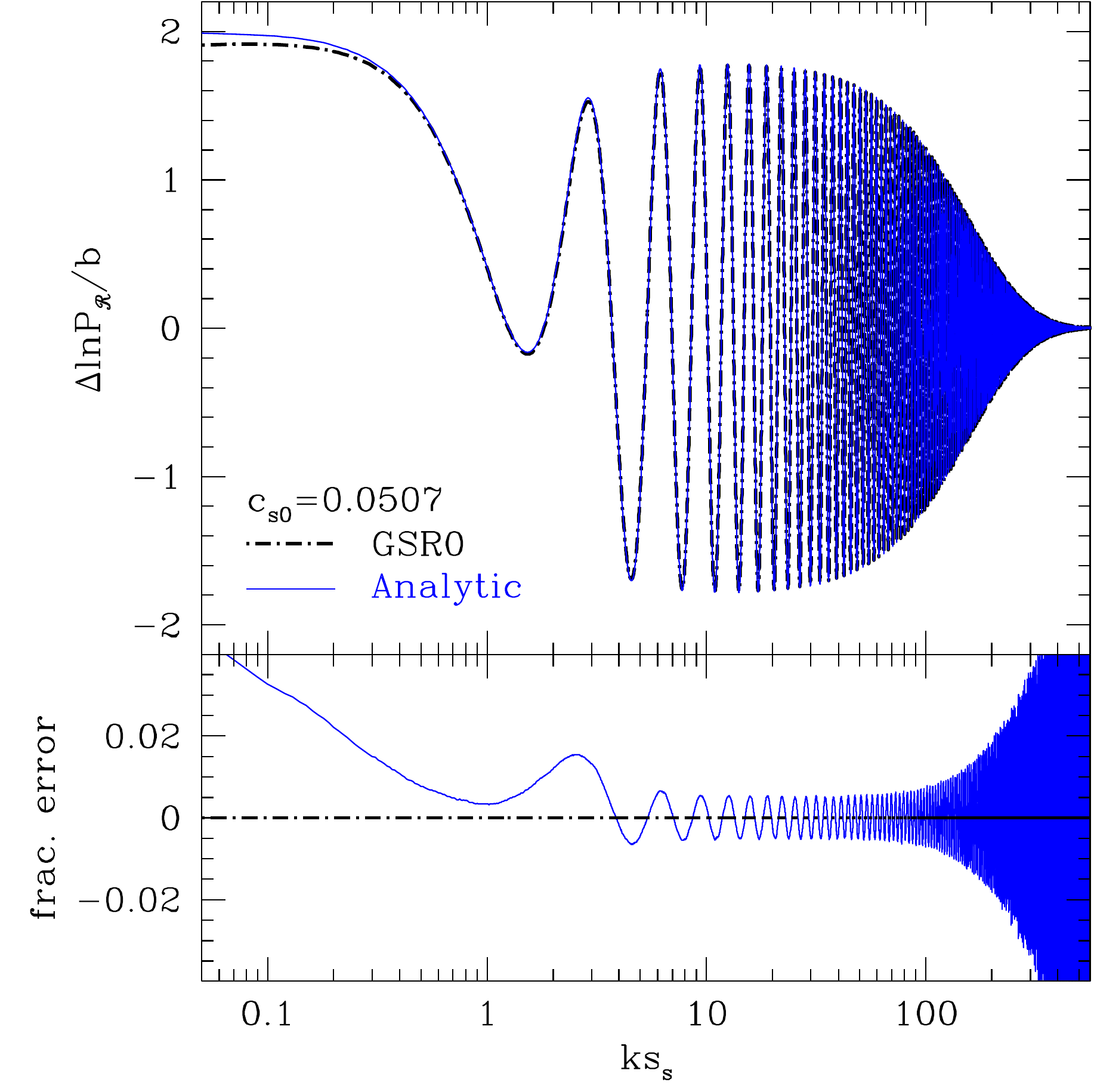, width=3.5in}
\psfig{file=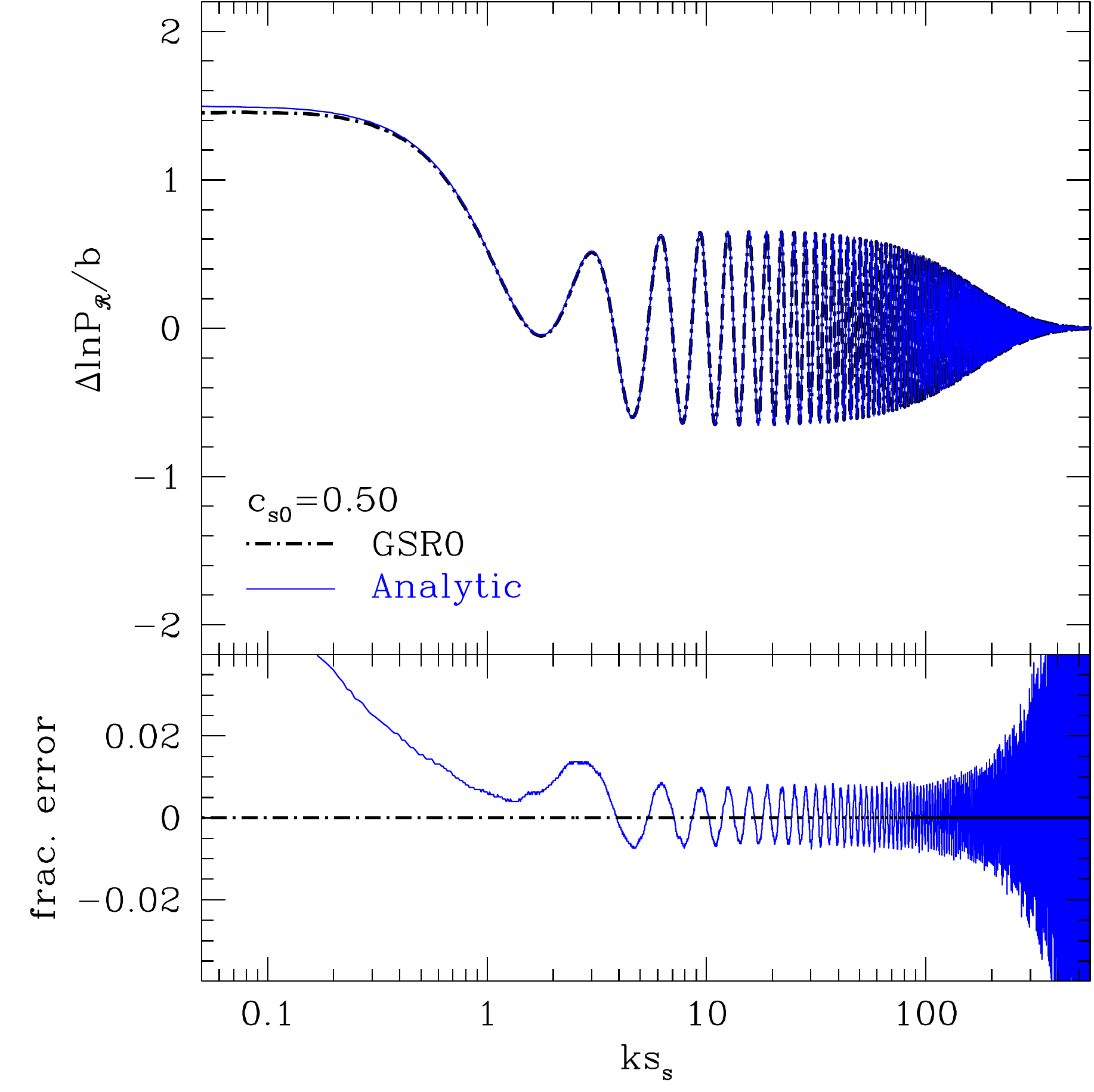, width=3.5in}
\caption{Analytic vs. GSR0 solution for a small amplitude sharp step $b=-0.005$,  $d = 0.005 \phi_{N0}=2.44 \times 10^{-12}$, with  $c_{s0}=0.0507$ (left) and $0.50$   (right).  Top panel: difference in 
$\ln P_{\curv}$ between this model and the same $b=0$ model.  Bottom panel:
difference between the curves in the top panel divided by the smooth envelope
of the oscillations (see text).
}
\label{fig:GSRvAnasmall}
\end{figure*}

        In this section, we will derive an analytical solution for  a very small and very sharp step in the warp factor and then test it against order unity steps.  
        From Fig.~\ref{fig:GSRvAnasmall}, we can infer that for $b\ll 1$ an analytic
        model based on integrating derivatives of $T(\phi)$ should be accurate once the appropriate conversions between the field and  sound horizon are made.  
        At larger $b$ we can see that the main differences are that the location,
        amplitude, and width of the feature in $G'$ changes which we shall see require a recalibration of corresponding effects in the power spectrum.
We use this analytic approximation to show that there is a DBI equivalent
to the sharp potential step model that improves
the WMAP7 likelihood by $2\Delta\ln {\cal L} \sim 12$ \cite{Adshead:2011jq}.

%\subsection{Analytic Approximation}

        We first start with some general
        considerations dictated by energy conservation and the slow-roll attractor.
If the inflaton crosses a step in $\delta N \ll 1$ then we can ignore energy loss to the expansion and set the total energy $\rho$ in Eq.~(\ref{eqn:energypressure}) to be equal before and immediately after the crossing \cite{Bean:2008na}.  Kinetic energy in excess (or deficit) of the  attractor after the step will then dilute away on the 
$\delta N \sim 1$ timescale.  
Denoting with $\Delta$ the change in quantities going through the step, we have
immediately after the step
\begin{equation}
\frac{\Delta c_s}{c_s}  = \frac{1-c_s}{1+ c_s \Delta T/T}\frac{\Delta T}{T}.
\end{equation}
Note that for a decrease in $T$, energy conservation restricts an amplitude
of $|\Delta T/T| = |2 b| < 1/c_s$.
For a small amplitude warp feature, we can linearize
\begin{equation}
\frac{\Delta c_s}{c_s}\approx (1-c_{s})\frac{\Delta T}{T} .
\label{eqn:csatstep}
\end{equation}
Thus for the case of a small, sharp step in $T$, the sound speed takes a fractional step of comparable amplitude.  Furthermore the slow-roll parameters
$\sigma_1$ and $\sigma_2$
follow by taking derivatives of $\Delta T/T$ during the interval around the step.
Similarly
\begin{equation}
\epsilon_H = \frac{3}{2} \frac{\rho+p}{\rho} \approx \frac{3}{2}\left( \frac{1}{c_s}-c_s \right) \frac{T}{V} ,
\end{equation}
and so
\begin{align}
\frac{\Delta \epsilon_H}{\epsilon_{H}} &=
\frac{c_s}{1+c_s \Delta T/T} \frac{1-c_s}{1+c_s}   \frac{\Delta T}{T} \nonumber\\
& %= c_{s} \frac{1-c_{s}}{1+c_{s}}\frac{\Delta T}{T}
= \frac{c_{s}}{1+c_{s}} \frac{\Delta c_s}{c_s}.
\label{eqn:epsatstep}
\end{align}
  Note that at low sound speed, the relative effect of the step on 
$\epsilon_H$ is suppressed vs $c_s$ by $c_s/(1+c_s)$, as are
$\eta_H$ compared with $\sigma_1$, and $\delta_2$ compared with 
$\sigma_2$, in agreement with Eq.~(\ref{eqn:etahsigma1}).

%\begin{figure}[t]
%\psfig{file=./plots/plot7.pdf, width=3.5in}
%\caption{GSR source function $G'$ for a warp step with $b=-0.05$ and $d=0.005\phi_{N0}$.   The source resembles the derivative of a delta-like function with width given by the step width $d$ and the speed of the inflaton $d\phi/ d\ln s$.   Also shown is the approximate form based on $T_{\phi\phi}$ in the sharp, small amplitude limit from Eq.~\eqref{eqn:appdG}.  }
%\label{fig:dG}
%\end{figure}

After crossing the step, we know that the inflaton hits the attractor solution (\ref{eqn:attractor})  as the
kinetic energy from the step decays after several efolds.
For a small
amplitude step
\begin{equation}
{\Delta c_{s} \over c_{s}}  = \frac{1}{2} (1-c_{s}^2) {\Delta T \over T},
\label{eqn:csafterstep}
\end{equation}
and 
\begin{equation}
\frac{\Delta \epsilon_{H}}{ \epsilon_{H}} ={\Delta c_s \over c_{s}} .
\label{eqn:epsafterstep}
\end{equation}
Given that the change in  $\epsilon_H$ is determined by the change in the sound speed
we seek to quantify the full evolution of $c_s$ from the step through to the attractor
regime.   

Following  \cite{Adshead:2011jq},
we begin by expanding the field as
\begin{equation}
\phi = \phi_0 + \phi_1,
\end{equation}
where $\phi= \phi_0$ when $b=0$,
and calculate to zeroth order in the unperturbed slow-roll parameters. Furthermore, the expansion rate is also unaffected by the warp feature since $\epsilon_H \ll 1$ throughout.
The field equation for $\phi_1$ then becomes
\begin{align}
\phi_{ NN 1} & =-  3 \phi_{N 1} + \frac{3}{2}(1-c_{s0}^2) \phi_{N 0 } \frac{T_1}{T_0} 
\nonumber\\
&\quad + \frac{1}{2} \frac{(1-c_{s0})^2(1+2 c_{s0})}{1-c_{s0}^2} \phi_{ N0 }^2 \frac{T_{\phi 1}}{T_0},
\end{align}
where again $0$ and $1$ denote unperturbed $b=0$ and finite $b$ perturbations
respectively.
Here we have used the fact that
\begin{align}\label{eqn:cs_1order}
    \frac{c_{s1}}{c_{s0}} & =   \frac{(c_{s0}^2 - 1)}{2 c_{s0}^2}  \left(2 \frac{\phi_{N1}}{\phi_{N0}} - \frac{T_{1}}{T_{0}} \right).
\end{align}
We can further transform the time variable from efolds $N$ to background field value $\phi_0$
by taking $\phi_{N0} \approx $ const.
\begin{align}
\frac{d}{d\phi_0} \left( e^{\frac{3\phi_0}{\phi_{N0}}} 
\frac{d\phi_1}{d\phi_0}\right) &= e^{\frac{3\phi_0}{\phi_{N0}}}  \Big[
\frac{3}{2}(1-c_{s0}^2) \frac{1}{\phi_{N0}} \frac{T_1}{T_0} 
\nonumber\\
&\quad + \frac{1}{2} \frac{(1-c_{s0})^2(1+2 c_{s0})}{1-c_{s0}^2}  \frac{T_{1\phi }}{T_0} \Big] .
\end{align}
The first term on the RHS can be integrated by parts to make the whole source proportional to
$T_{1\phi}$.
For sharp features, $T_{1\phi}$ is very concentrated around the feature and, consequently, we can approximate the background quantities by their values at $\phi_s$.  Combined with
the boundary condition that the field is on the attractor before the step
\begin{equation}
\lim_{\phi \ll \phi_s} \frac{d\phi_1}{d\phi_0} = -(1-c_{s0}^2) b,
\end{equation}
we obtain
\begin{align}
	 \frac{d\phi_{1}}{d\phi_{0}}  &=
	\frac{1- c_{s0}^2}{2} b F(\phi_0) -\frac{c_{s0}^2}{2}  \frac{1-c_{s0}}{1+c_{s0}} b[F(\phi_0)+2]e^{\frac{3 (\phi_{s} - \phi_0)}{\phi_{N0}}}.
	  \end{align}
Using this result in Eq.~\eqref{eqn:cs_1order} and replacing
\begin{equation}
\phi_s-\phi_0 = \phi_{N0} (N_s-N),
\end{equation}
we obtain
\begin{align}
	 \frac{c_{s1}}{c_{s0}} =  \frac{1-c_{s0}^2}{2} b F(\phi_0) + 
	 \frac{\left(1 - c_{s0}\right)^2}{2}    b [F(\phi_0)+2] e^{{3(N_s-N)}}.
\end{align}
Note that before the step $c_{s1}/c_{s0} = -(1-c_{s0}^2)b$ and right after the
step $c_{s1}/c_{s0} =  (1-c_{s0})^2 b$ and so $\Delta c_s/c_s = 2(1-c_{s0}) b$
as expected from  Eq.~(\ref{eqn:csatstep}).  Several efolds after the step 
$c_{s1}/c_{s0} = 0$ and so  $\Delta c_s/c_s = (1-c_{s0}^2)b$
as expected from Eq.~(\ref{eqn:csafterstep}). 

From $\epsilon_H = \phi_N^2/2c_s$ we obtain
\begin{align}
\frac{ \epsilon_{H1}}{\epsilon_{H0}} &= \frac{1- c_{s0}^2}{2} b F(\phi_0)  \\
&\quad - \frac{1+c_{s0}^2}{2} \frac{(1-c_{s0})}{(1+c_{s0})}     b [F(\phi_0)+2] e^{{3(N_s-N)}} ,\nonumber
\end{align}
%
%\begin{align}
%\frac{ \epsilon_{H1}}{\epsilon_{H0}} &= \frac{1}{2}
%\left[ (1+ c_{s0})^2 - (1+ c_{s0}^2) e^{{3(N_s-N)}}\right] \nonumber\\
%&\quad \times \frac{(1-c_{s0})}{(1+c_{s0})}  b F\left(\frac{\phi_{0} - \phi_s}{d} \right) 
%\end{align}
which also satisfies expectations from Eqs.~(\ref{eqn:epsatstep})
and (\ref{eqn:epsafterstep}).

\begin{figure*}[t]
\psfig{file=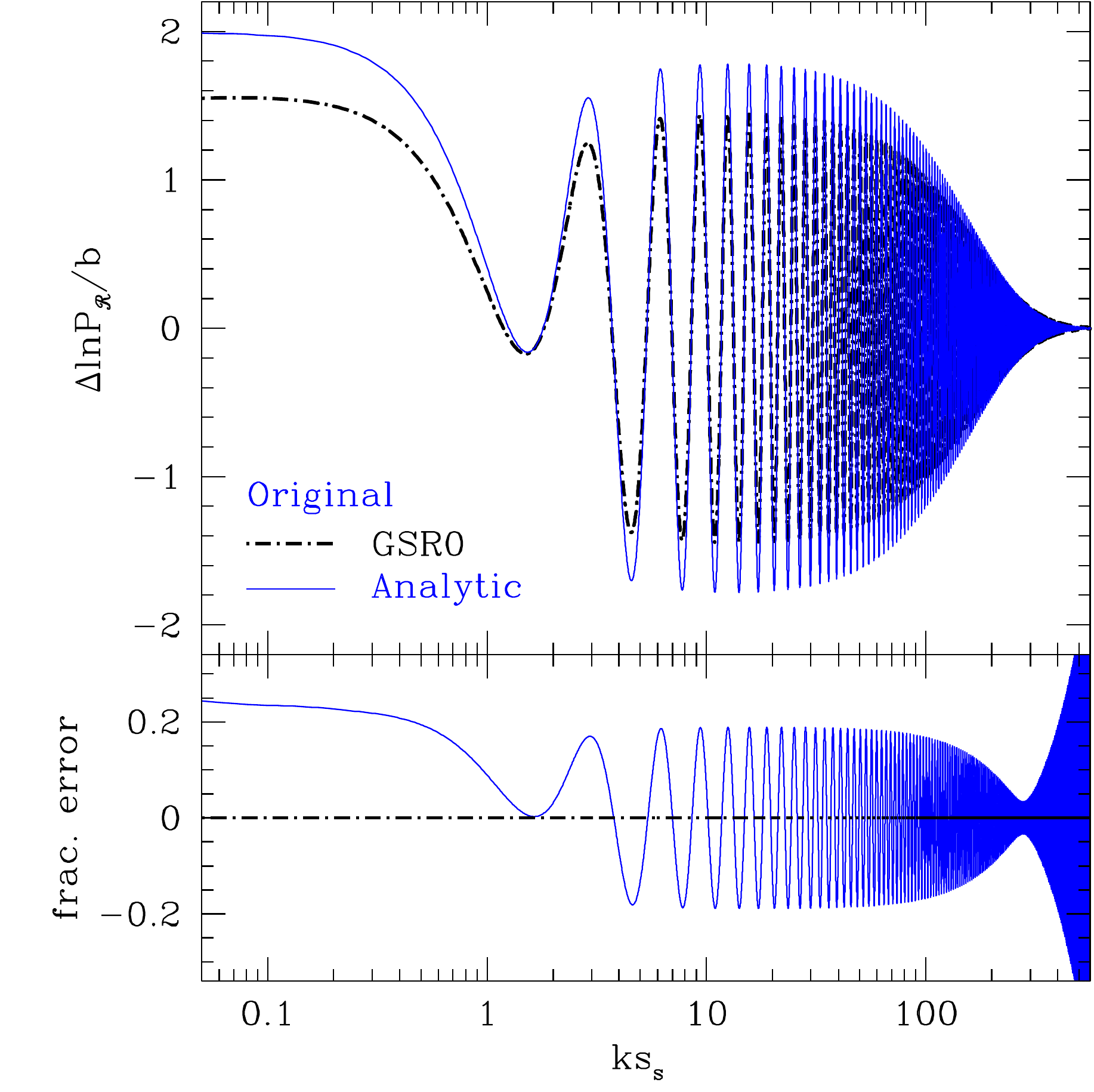, width=3.5in}
\psfig{file=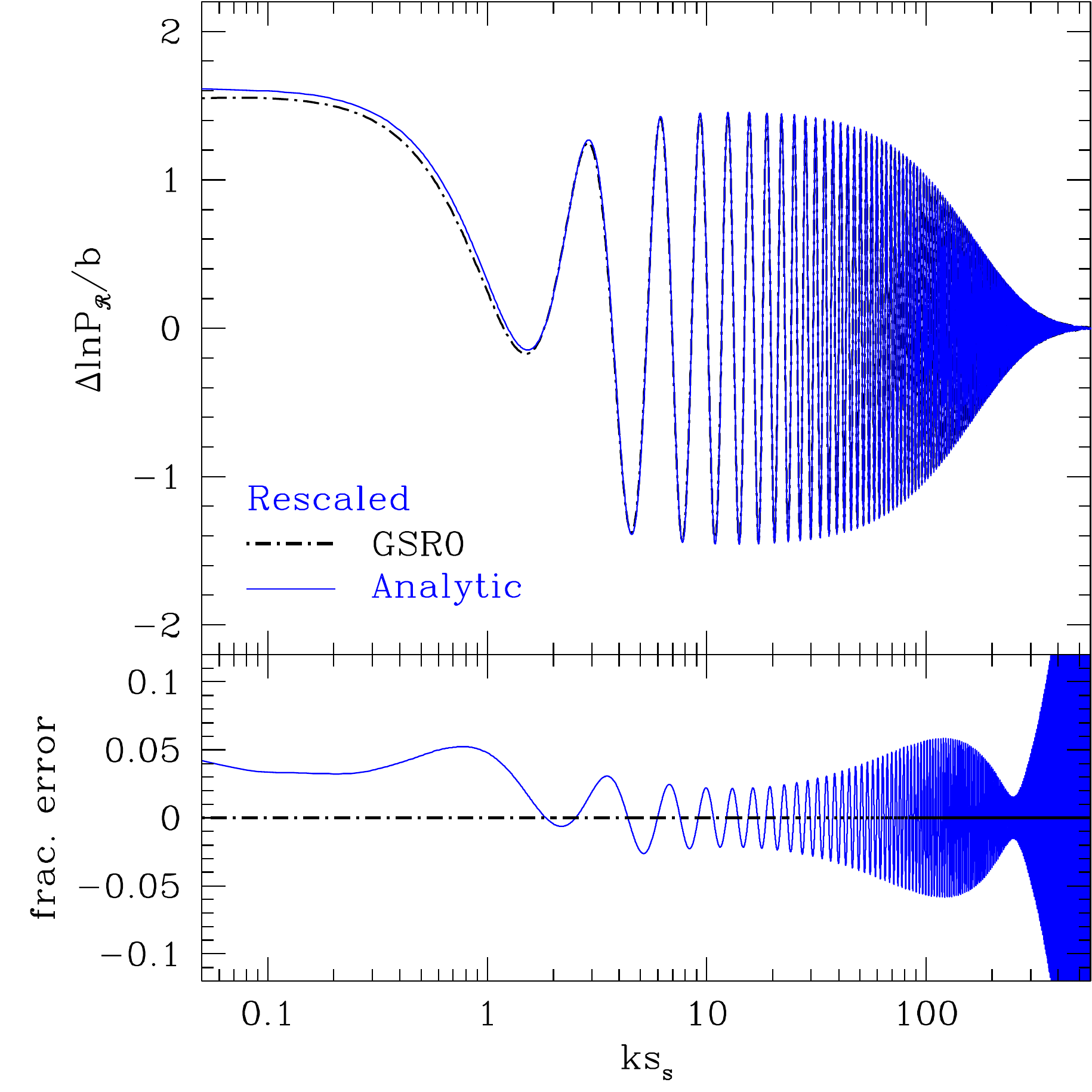, width=3.5in}
\caption{Analytic vs. GSR0 solution for a large amplitude sharp step $b=-0.25$ and $d = 2.44 \times 10^{-12}$ with low sound speed $c_{s0}=0.0507$ model as in Fig.~\ref{fig:GSRvAnasmall}.  Left: the  analytic approximation using the linear in $b$ scalings of Eq.~(\ref{eqn:Cfactors}) begins to depart from GSR0 as $b$ approaches unity.   Note however that the functional form of the feature
remains the same but amplitudes and damping require rescaling. Right: nonlinearly rescaled amplitudes and damping of
Eq.~(\ref{eqn:Cfactorsrescaled}) recover the few percent level accuracy seen in the
small step case. }
\label{fig:GSRvAnalarge}
\end{figure*}

From these quantities, we calculate $G'$ taking $\epsilon_H \rightarrow 0$,
$b\rightarrow 0$, $d\rightarrow 0$
\begin{eqnarray}
G'& \approx & -\frac{1}{3}\sigma_2 + \frac{2}{3} \delta_2  -\frac{5}{3} \sigma_1
- 2\eta_H  +  {8 \over 3} ({ a H s \over c_s} -1).
\end{eqnarray}
In this approximation,
\begin{align} 
\frac{a Hs}{c_s}-1&\approx  -\frac{5-2 c_{s0} - 3 c_{s0}^2}{8}b F(\phi_0)e^{{N-N_s}}
\nonumber\\
&\quad  - \frac{3(1-c_{s0})^2}{8}
  b [F(\phi_0)+2] e^{{3(N_s-N)}} .
\end{align}
After several integrations by parts  we obtain the change in $\ln \Delta_{\curv}^2$
from Eq.~(\ref{eqn:GSRpower}) due to the feature, with $I_1=I_2=0$ from the feature, as 
 \begin{align}
\ln \Delta_{\curv 1}^2 &=   C_1 W(ks_s)  +C_2  W'(ks_s)+ C_3 Y(ks_s),%+\frac{1}{4}(1-c_{s0}^2)Y_2(k s_s),
\label{eqn:powerform}
\end{align}
where
\begin{align}
Y(x) &= \frac{6 x \cos(2 x) + (4x^2 -3) \sin(2 x)}{x^3} %\\
%Y_2(x) &= -\frac{14 x \cos(2 x) + (4x^2 -19) \sin(2 x) + 12 {\rm Si}(2 x)}{x^3}. \nonumber
\end{align}
%\wh{This derivation needs to be checked thoroughly - the $Y_2$ term looks like it may have
%too large an amplitude to agree with numerics}
is proportional to $\int d\ln x W'/x$.   Here
\begin{align}
C_1 &= 2 (1-c_{s0}^2)b ,\nonumber\\
C_2 &= -\frac{2}{3}  \frac{1-c_{s0}}{1+c_{s0}}b,\nonumber\\
C_3 & =  \frac{5-2 c_{s0} - 3 c_{s0}^2}{4}b.
\label{eqn:Cfactors}
\end{align}
Given that
\begin{align}
\lim_{x\rightarrow 0} W(x) & = 1, \quad 
\lim_{x\rightarrow \infty} W(x) = 0, 
\nonumber\\
\lim_{x\rightarrow 0} W'(x) & =0, 
\quad \lim_{x\rightarrow \infty} W'(x) = -3\cos(2x),
\nonumber\\
\lim_{x\rightarrow 0} Y(x) & =0, 
\quad \lim_{x\rightarrow \infty} Y(x) =0,
\end{align}
 the $W$ term represents a step in the power spectrum at $k s_s\sim 1$ of fractional 
amplitude $2b(1-c_{s0}^2)$, which follows directly from the attractor solution, and
the $W'$ term represents a constant amplitude oscillation at $k s_s \gg 1$.  The latter is exactly the same form as oscillations produced by a step in the potential
for a canonical kinetic term (see~\cite{Adshead:2011jq} Eq.~32).  Unlike the canonical case,
the step in power is comparable to the amplitude of oscillations.
Furthermore, the additional $Y$ term changes the solution near $k s_s \sim 1$.
Since even for $b\ll 1$ a small error in the location of the feature $s_s$, which controls the frequency of the oscillation, causes a noticeable change in the phase of the oscillation over many cycles, we define
it such that 
\begin{equation}
\delta G'(\ln s_s)=0
\end{equation}
 for the change from the smooth $b=0$ model.
This definition differs slightly from the sound horizon at $\phi_s$ for large $b$
as shown in Fig.~\ref{fig:dG}.

For finite step width $d$ in field space, the inflaton traverses the step in 
$\Delta s/s_s \approx  |d\ln s/d\phi| d \approx d/\phi_{N0}$.   The window functions $W$ and $W'$ oscillate on a time scale $\Delta s=1/k$.   Thus the integral
over $G'$ is damped for $k s_s > \phi_N/d$.  For the tanh step, the integral can be approximated following \cite{Adshead:2011jq}
 \begin{align}
\ln \Delta_{\curv 1}^2 &= \Big[ C_1 W(ks_s)  + C_2 W'(ks_s) 
%\nonumber\\&\quad
 +C_3  Y(ks_s) %+\frac{1}{4}(1-c_{s0}^2)Y_2(k s_s)
\Big]
\nonumber\\&\quad \times 
{\cal D} \left( \frac{k s_s}{x_d} \right),
\end{align}
where
\begin{equation}
x_d = \frac{ d\phi}{d \ln s} \frac{1}{\pi d} \approx \frac{\phi_{N0}}{\pi d} = \frac{\sqrt{2\epsilon_{H0}c_{s0}}}{\pi d},
\end{equation}
and the damping function is
\begin{equation}
{\cal D}(y) =  \frac{y}{\sinh y}.
\end{equation}
To obtain the full power spectrum we add $\ln \Delta_{\curv 1}^2$ to a calculation of 
the $b=0$ model.   This can be an exact numerical solution, a slow-roll approximation, or
the GSR approximation.   For comparison purposes, we choose here to take the GSR0  ($I_1=I_2=0$) solution
from  \S \ref{sec:GSR}.

In Fig.~\ref{fig:GSRvAnasmall}, we test the analytic approximation for a small amplitude sharp step  $b =- 0.005$ and two values of the sound speed.    In the lower panel we divide the difference
between the analytic and  GSR0 solutions  by the envelope function
\begin{equation}
3 C_2 {\cal D}\left( \frac{k s_s}{x_d} \right).
\end{equation}
  Agreement is at the 1\% level except on
scales much larger than the step $k s_s \ll 1$ and those affected by damping  $k s_s \gtrsim x_d$.  In the former case differences from the change in $\phi_N$ from $\phi_{N0}$ due to the different slow-roll attractor change the mapping between
$\phi$ and $\ln s$.   Near the damping scale, small changes in $x_d$ are amplified in the fractional difference due to the exponential nature of the damping even though the absolute prediction remains accurate.

\begin{figure}[t]
\psfig{file=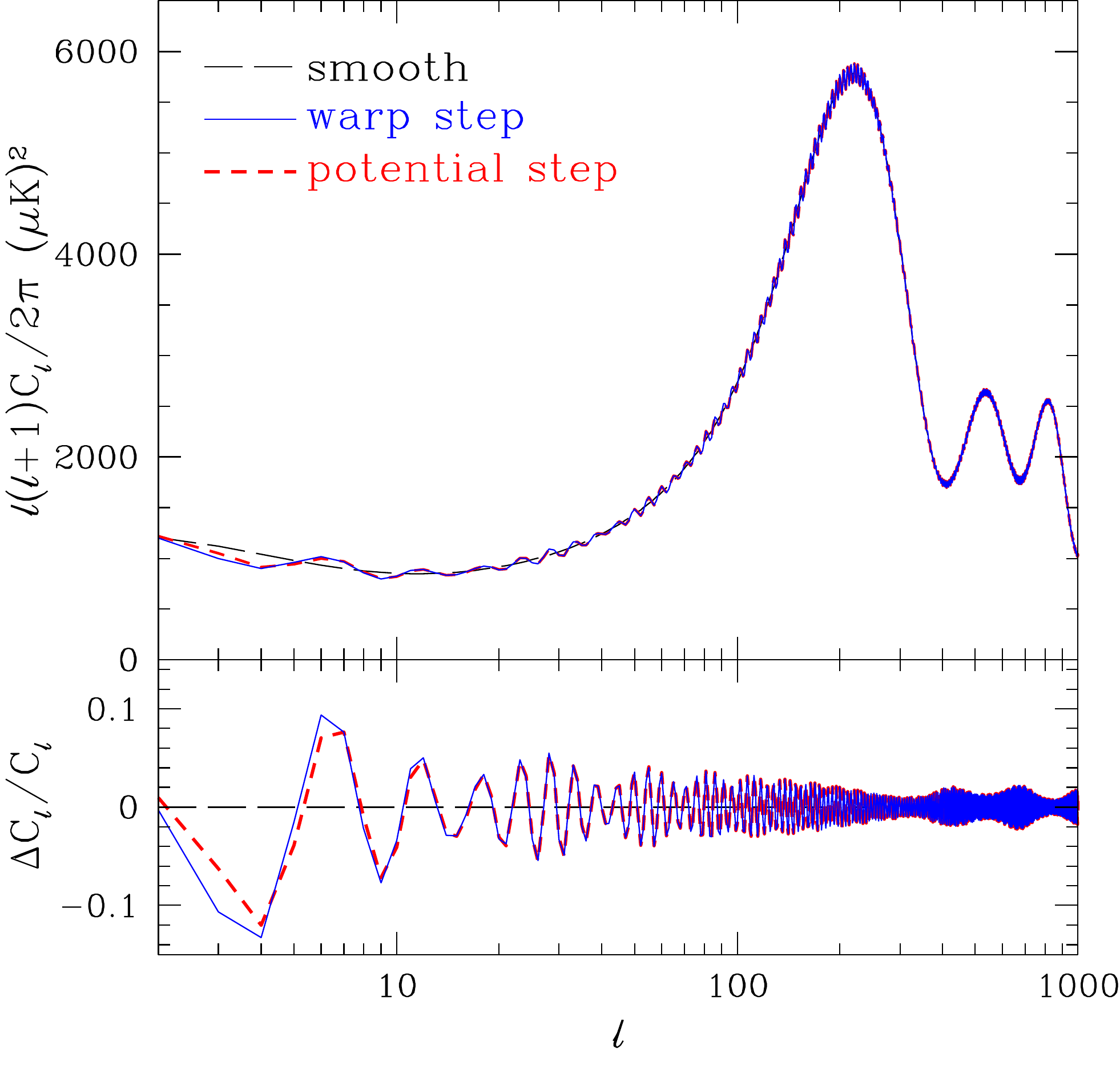, width=3.25in}
\caption{CMB temperature power spectrum for the a horizon scale, sharp step in the potential $V(\phi)$ of a canonical inflation fit to the WMAP7 data versus the matching step in the DBI warp $T(\phi)$.  Lower panel shows the fractional deviation from a smooth spectrum with the 
same average power.
 The two types of steps produce nearly identical deviations at high $\ell$ and so fit the  data equally well.  Both spectra
are calculating using their respective analytic approximations and without gravitational lensing.}
\label{fig:cl}
\end{figure}

As $b$ approaches order unity, the analytic approximation 
begins to misestimate the amplitudes and damping of the features.
In Fig.~\ref{fig:GSRvAnalarge} (left) we show a case where $b=-0.25$ and
$d=2.44 \times 10^{-12}$, with low sound speed $c_{s0}=0.0507$.
Note that the functional form of the power spectrum
in Eq.~(\ref{eqn:powerform}) remains the same only the coefficients differ.
We therefore rescale them as
\begin{align}
C_1 &=  \ln\left[ \frac{1-2 b c_{s0}^2}{1-2b} \right], \nonumber\\
C_2 &=  -\frac{2}{3}  \frac{1-c_{s0}}{1+c_{s0}}\frac{1}{\sqrt{1-2b}}b ,\nonumber\\
C_3 &=  \frac{5-2 c_{s0} - 3 c_{s0}^2}{4} \frac{1}{\sqrt{1-2b}}b, \nonumber\\
x_d & = \frac{d\phi}{d\ln s}\Big|_{s_s} \frac{1}{\pi d},
%x_d &= \frac{\sqrt{2\epsilon_{H0}c_{s0}}}{\pi d} \left[ 1-  
%\frac{b}{2}\frac{1+ c_{s0}- 2 c_{s0}^3}{1+c_{s0}}\right].
\label{eqn:Cfactorsrescaled}
\end{align}
where the form of $C_1$ can be derived from the attractor solution and the form of the $C_2$, $C_3$ corrections is motivated by the
fact that $T=0$ before the feature for $b=1/2$.   
In Fig.~\ref{fig:GSRvAnalarge} (right), we show that the agreement is again good 
after these rescalings even for the $b=-0.25$ case.

Given that the analytic approximation works quite well even for relatively large
values of $b$ and its functional form mimics a step in the potential at $k s_s \gg 1$, we can remap results for the latter onto the former.    A potential step
at $s_s = 8163$ Mpc of amplitude $C_2=0.11$ improves the WMAP7 
likelihood by $2\Delta \ln {\cal L} \approx 12$ \cite{Adshead:2011jq}.  
In terms of the warp step, these parameters translate into
$b=-0.218$ for $c_{s0}=0.0507$ and $d \rightarrow 0$.   In Fig.~\ref{fig:cl} we compare the CMB temperature
power spectra predicted by the two models using the best fit parameters for the cosmological parameters: $\Omega_b h^2=0.0222$, $\Omega_c h^2=0.11$, $h=0.71$, $\tau=0.10$,
$\ln 10^{10} A_s =3.077$, $n_s=0.965$ such that the underlying smooth power spectrum
is
\begin{equation}
\Delta_{{\cal R}0}^2= A_s \left( \frac{k}{0.05{\rm Mpc}^{-1}} \right)^{n_s-1}.
\end{equation}   Note that aside from small changes
at low multiple $\ell$ where the cosmic variance is high, the two spectra are
indistinguishable.   Thus a step in the warp fits the WMAP7 data as well
as a step in the potential.

\section{Discussion}
\label{sec:discussion}

We have shown that the GSR approximation can be applied to DBI inflation to constrain features in the warped brane tension $T(\phi)$ from observational data.   The approximation accurately recovers corresponding features for up to order unity 
deviations.  Previous work on constraining the GSR source function $G'$ and hence
second derivatives of the potential $V(\phi)$ for canonical fields can 
be directly reinterpreted in the DBI context as limits on the second derivative
of $T(\phi)$ \cite{Dvorkin:2010dn,Dvorkin:2011ui}.  The main difference between the two is that features in $T(\phi)$ once traversed can strongly affect the slow-roll
attractor for modes that cross the sound horizon later.

The correspondence between features in $V(\phi)$ and $T(\phi)$ is especially
close in the limit of extremely sharp features, for example a step feature.  In both cases the power spectrum
exhibits constant amplitude oscillations for modes that cross the sound horizon after the step.  Consequently, the preference for a horizon-sized step in the potential in WMAP7 implies a corresponding preference for a step in $T(\phi)$.
The main difference is a reduction of power for the low $k$ modes that cross before the feature.   The large cosmic variance of these modes prevents a
significant distinction between the two.   On the other hand,  features in $V(\phi)$ for canonical inflation
and $T(\phi)$ for DBI inflation should induce very different bispectra.   We leave
these considerations to a future work.

\acknowledgements
We thank Rachel Bean, Xingang Chen, Cora Dvorkin, Hiranya Peiris and Mark Wyman for useful discussions. VM was supported by the Brazilian Research Agency CAPES Foundation and by  U.S. Fulbright Organization. VM and WH were  supported in part by U.S.~Dept.\ of Energy contract DE-FG02-90ER-40560.   PA and WH were supported  by the Kavli Institute for Cosmological Physics at the University of Chicago through grants NSF PHY-0114422 and NSF PHY-0551142 and an endowment from the Kavli Foundation and its founder Fred Kavli. 
WH was additionally supported by  the David and Lucile Packard Foundation.
\vfill

\bibliography{DBI}

\end{document}